# Structure, hydrogen bond dynamics and phase transition in a prototypical ionic liquid electrolyte.

Alexander E. Khudozhitkov[a,d], Peter Stange[b], Alexander G. Stepanov[a], Daniil I. Kolokolov[a,d]*, Ralf Ludwig[b,c,]*

a  Boreskov Institute of Catalysis, Siberian Branch of Russian Academy of Sciences, Prospekt Akademika Lavrentieva 5, Novosibirsk 630090, Russia; E-mail: kdi@catalysis.ru

b  Universität Rostock, Institut für Chemie, Abteilung für Physikalische Chemie, Dr.-Lorenz-Weg 2, 18059 Rostock, Germany; Tel: 49 381 498 6517; E-mail: ralf.ludwig@uni-rostock.de

c  Leibniz-Institut für Katalyse an der Universität Rostock e.V., Albert-Einstein-Str. 29a, 18059 Rostock (Germany)

d  Novosibirsk State University, Pirogova Street 2, Novosibirsk 630090, Russia

### Abstract

Ionic liquids (ILs) gain much interest as possible electrolytes in the next generation of mixed-solid Li-ion batteries. However, such properties of ionic liquids as melting transition, diffusion, strength and structure of hydrogen bond network remain poorly investigated. Here $^2$H NMR study of a prototypical ionic liquid [TEA][NTf$_2$] has been performed. We show that the dynamical melting occurs through a dynamically heterogeneous phase stable between 223-277 K and the transition process is characterized by two standard molar enthalpy changes indicating to the multistage nature of the melting. The spin relaxation analysis allowed determining geometry, rates and energetics of IL mobility in both solid and liquid state. We compare these properties with previously reported data on [TEA][OTf] and [TEA][OMs] ILs that share cation but have anions of varying strength. Our results prove that the stronger hydrogen bonds between cation and anion lead to the lower enthalpy change between solid and liquid state, higher activation barrier of tumbling motion and lower amplitude of libration motion.

# 1. Introduction.

Bis(trifluoromethanesulfonyl)imide (NTf$_2$) is the most frequently used anion due to its ability to form only weak hydrogen bonds within the cation-anion pair, thus creating opportunity for the cations to migrate on its own and not only within an ionic pair. This is was found to be particular important in electrochemical applications as it greatly decrease the viscosity of the ionic liquid (IL) which is beneficial for the ionic transport. Currently the ionic liquids tested as potential electrolytes are predominantly based on the NTf$_2$ anion. In fact, protic ionic liquids (PILs) based on the 3-alkyl ammonium cation, such as [3-ethylammonium] [Bis(trifluoromethanesulfonyl)imide] ([TEA][NTf$_2$]), were recently shown to provide good transport and stability properties for the Li ions.[1] In the same time the knowledge about the microscopic diffusion in similar PILs remains often unknown even for the model electrolytes such as the [TEA][NTf$_2$]. Moreover, such properties as the structure of hydrogen bonds in solid and liquid phase, their mobility and phase behaviour remain poorly characterized and understood. These questions are currently of pivotal interest, as more these materials are envisioned as key components of the next generation of mixed-solid Li-ion batteries capable to operate in a broad temperature range.

Recently we have developed a robust approach to solve the problem in liquids and frozen PIL by using the solid state $^2$H NMR line shape and spin relaxation analysis.[2, 3]. In particular, we were able to show the presence a dynamical heterogeneous phase in PILs based on the [TEA]$^+$ anion. Moreover, measurement of the temperature-dependent spin-lattice relaxation time allowed to probe dynamics beyond to the liquid phase, covering both the heterogeneous and the low temperature solid phase. With this in mind, we decided to investigate the mobility and structure of the hydrogen bonds in the [TEA][NTf$_2$] protic ionic liquid.

In this report we provide a detailed solid state $^2$H NMR investigation of the [TEA][NTf$_2$] protic ionic liquid selectively labelled in its hydrogen, i.e. [(C$_2$H$_5$)$_3$ND][NTf$_2$], over a broad temperature range of 133-423 K.

# 2. Experimental section

## 2.1. Materials

The synthesis of triethylammonium bis(trifluoromethane-sulfonyl)imide material [(C$_2$H$_5$)$_3$ND][NTf$_2$] was performed according to the previously reported procedure.[4] Hydrogen/deuterium (H/D) exchange was achieved by mixing the PILs with D$_2$O and removing water several times until nearly 100% deuteration was reached as proven by $^1$H NMR. All samples have been dried under vacuum (at 3·10⁻³ mbar) for several days and the final water concentration (< 15 ppm) has been checked by Karl-Fischer titration.

### *2.2. Sample preparation*

Sample preparation for the NMR experiments was performed in the following manner. The ionic liquid was loaded into a glass tube (5 mm outer diameter; 20 mm long), connected to a high vacuum grade valve (HI-VAC). All manipulations were performed in argon atmosphere. The sample was then attached to a vacuum line and the argon was pumped off under vacuum to a final pressure above the sample of $10^{-2}$ Pa. To fully degas the material the sample was slowly introduced into liquid nitrogen 2-3 times, while being connected to vacuum line. After degassing, the neck of the tube was sealed off, while the sample material was maintained in liquid nitrogen in order to prevent its heating by the flame. The sealed sample was then transferred into an NMR probe for analysis with $^2$H NMR spectroscopy.

### *2.3. $^2$H NMR experiments*

$^2$H NMR experiments were performed at the Larmor frequency $\omega_0/2\pi = 61.424$ MHz on a Bruker Avance-400 spectrometer using a high-power probe with a 5 mm horizontal solenoid coil. All $^2$H NMR spectra were obtained by a Fourier transform of a quadrature-detected and phase-cycled quadrupole echo after two phase-alternating 90°-pulses in the pulse sequence ($90°_x - \tau - 90°_y - \tau -$ acquisition$-$ t), where $\tau = 20$ μs and t is a repetition time of the sequence during accumulation of the NMR signal.[5] The duration of π/2 pulse was 1.8-1.9 μs. Spectra were typically obtained with 400−2000 scans at low temperature and 40-80 scans at high temperature with a repetition time ranging from 0.2 to 3 s. Inversion−recovery experiments for measurements of spin−lattice relaxation times ($T_1$) were carried out using the pulse sequence $180°_x - t_v - 90°_x - \tau - 90°_y - \tau -$ acquisition $- t$, where $t_v$ was the variable delay between the 180° and the 90°-pulses. Spin-spin relaxation time ($T_2$) was measured by a Carr-Purcell-Meiboom-Gill pulse sequence.[6] The repetition time $t$ was always longer than 5-fold of the estimated relaxation time $T_1$.

The temperature of the samples was controlled with a flow of nitrogen gas by a variable-temperature unit BVT-3000 with a precision of about 1 K. The sample was allowed to equilibrate at least 15 min at the temperature of the experiment before the start of the NMR signal acquisition.

### *2.4. Modelling of NMR line shape and relaxation*

Modelling of the $^2$H NMR spectra line shape and spin relaxation times was performed with a homemade Fortran program based on the standard formalism applied for the description of molecular motions.[7, 8]

### 3. Results and discussion

$^2$H NMR spectra were collected in the 143–436 K temperature region. At low temperature spectrum has a Pake-like line shape (Fig. 1). Closer consideration reveals that spectra are in fact a superposition of two Pake-doublets with deuteron quadrupole coupling constants (DQCC) $\chi_{DIa}$ = 173 – 176 kHz, $\chi_{DIb}$ = 148 – 150 kHz and asymmetry parameter $\eta$ = 0 – 0.05. Coexistence of two signals implies that there are two distinct kinds of N-D groups in solid state of IL. The DQCC is determined by the magnitude of electric field gradient (*efg*) on the nucleus position. This *efg* originates mainly from the electrons of N-D group that participates in hydrogen bond formation. It is known from molecular liquids that the DQCC is a sensitive probe for hydrogen bonding and that DQCC value significantly increases from the solid via the liquid to the gas phase.[9-13] The presence of two kinds of N-D groups is explained by the possibility of the formation of two kinds of hydrogen bonds that was supported with density function theory (DFT) calculation.[2]

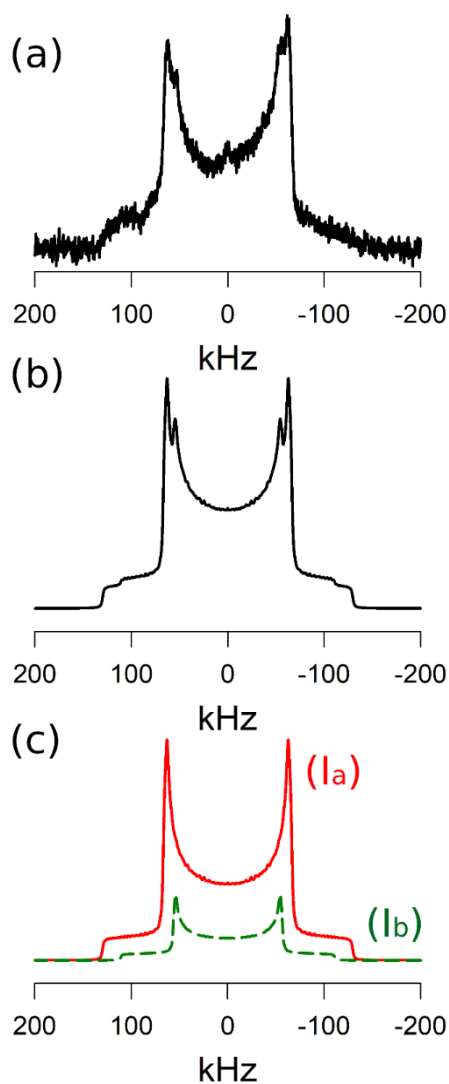

**Figure 1.** $^2$H NMR spectrum of [TEA][NTf$_2$] at 223 K (a) experimental, (b) simulated, (c) deconvolution of the spectrum.

At 223 K there appears narrow Lorentzian signal in the centre of $^2$H NMR spectrum that corresponds to isotropic liquid phase (Fig. 2). Upon heating the intensity of the isotropic signal increases and at 283 K only the isotropic signal remains (Fig. 3). This temperature is in a good accordance with the melting temperature 277 K derived with differential scanning calorimetry (DSC).[14] Noteworthy, that the samples were measured from lower to higher temperatures. In order to achieve fast cooling samples were quenched in liquid nitrogen. The measurements were reproducible after several days. Thus, we are sure that the samples were in equilibrium.

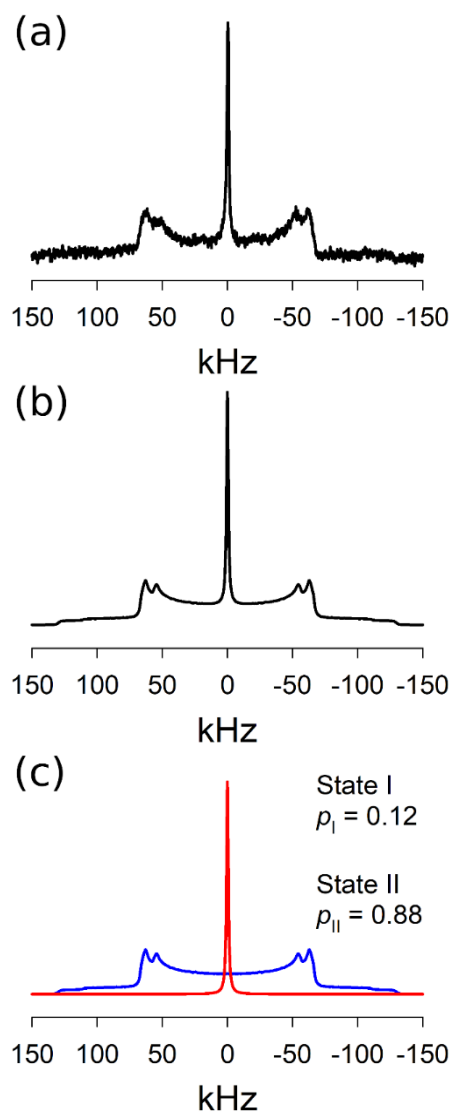

**Figure 2.** $^2$H NMR spectrum of [TEA][NTf$_2$] at 243 K (a) experimental, (b) simulated, (c) deconvolution of the spectrum.

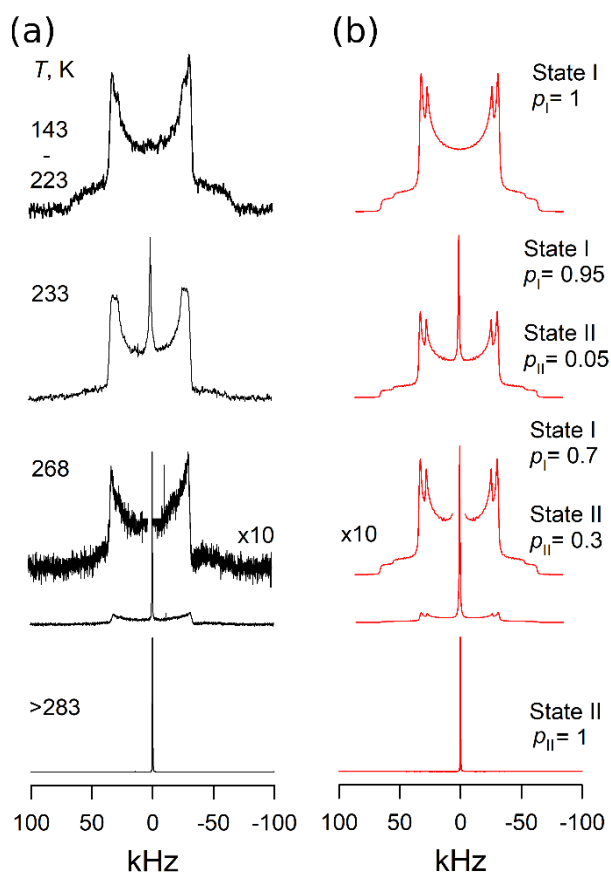

**Figure 3.** Temperature evolution of $^2$H NMR spectra of [TEA][NTf$_2$]: (a) experimental and (b) simulated spectra.

From the analysis of $^2$H NMR spectra line shape we can deduce the thermodynamic properties of transition from the rigid to mobile state of IL. Here we introduce the relative population of rigid [TEA]$^+$ cations $p_I$ (which includes both anisotropic components I$_a$ and I$_b$) and mobile fraction $p_{II}$ that are proportional to the relative intensity of isotropic and anisotropic components of $^2$H NMR spectrum respectively. The calculated isotropic and anisotropic fractions of the $^2$H spectra are shown in Fig. 4. The ratio gives the equilibrium constant $K = p_{II}/p_I$ between the isotropic and anisotropic fractions as a function of temperature, resulting in a van 't Hoff plot as shown in Fig. 5:

$$\ln(K) = -\frac{\Delta H^\ominus}{RT} + \frac{\Delta S^\ominus}{R}$$

(1)

Two standard molar enthalpy changes $\Delta H^\ominus$ can be taken from the slopes of the curve. For the initial process below 238 K $\Delta H_1^\ominus = 110 \pm 20$ kJ mol$^{-1}$, close to the second melting transition we obtained $\Delta H_2^\ominus = 22 \pm 5$ kJ mol$^{-1}$, indicating that less energy is required to reach the second melting transition. These melting transitions are more energy demanding compared to [TEA][OTf] ($\Delta H_1^\ominus = 67$ kJ mol$^{-1}$ and $\Delta H_2^\ominus = 17$ kJ mol$^{-1}$) but surprisingly similar to the case of the [TEA][OMs] ($\Delta H_1^\ominus = 124$ kJ mol$^{-1}$ and $\Delta H_2^\ominus = 20$ kJ mol$^{-1}$) characterized by a narrower temperature range of coexistence of the two dynamical states.[3, 15] Such relation between enthalpy changes is unexpected, since [NTf$_2$]$^-$ anion is known to interact weaker compared to [OTf]$^-$ and much weaker that the [OMs]$^-$. The latter is usually assumed to have a strong pairing, i.e. the ion pair rotates as a whole. In such case we believe that the high enthalpy change results from a more developed interpenetrating network of hydrogen bonds, which restrict the overall mobility within the solid PIL.

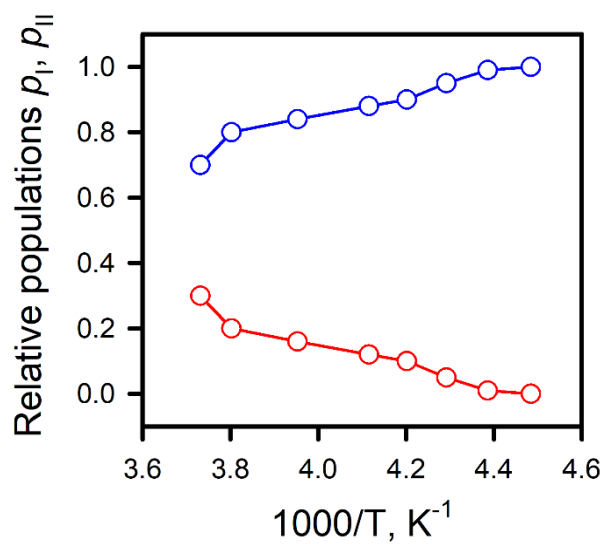

**Figure 4.** The relative populations of the anisotropic (blue) and isotropic (red) fraction of the $^2$H patterns of the PIL [TEA][NTf$_2$]. In the covered temperature range between 233 and 273 K both components are present.

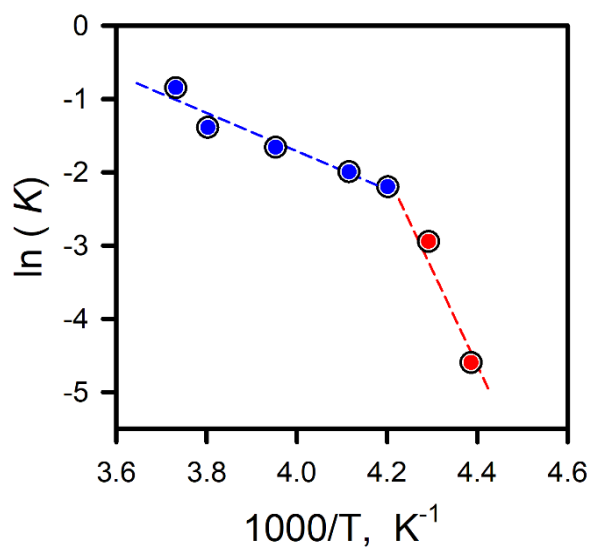

**Figure 5.** Van 't Hoff plot for the equilibrium constants $K$ obtained from the relative populations $p_{II}/p_I$ taken from the decomposed $^2$H patterns in Figure 4 between 233 K and 273 K. The dotted lines represent linear fits (R$^2 \geq 0.95$) with a slope of $\Delta H^\circ/R$.

In order to probe motional behavior of [TEA][NTf$_2$] we measured the deuteron spin-lattice $(T_1)_D$ and spin-spin $(T_2)_D$ relaxation times of the N–D molecular vector for the isotropic and anisotropic components. Deuteron nuclear magnetic relaxation is driven by the interaction of the electrostatic quadrupole moment $eQ$ of the deuteron nucleus with the main component of the *efg* tensor at the nucleus, $eq_{zz}$, generated by the electron distribution surrounding the nucleus along the N–D bond. The relaxation rates are given by [5-7, 16-21]

$$\left(\frac{1}{T_1}\right)_D = \frac{3}{20}\pi^2 \left(\frac{eQeq_{zz}}{h}\right)_D^2 \{J_1(\omega_0) + 4J_2(2\omega_0)\} \qquad (2)$$

$$\left(\frac{1}{T_2}\right)_D = \frac{1}{40}\pi^2 \left(\frac{eQeq_{zz}}{h}\right)_D^2 \{9J_0(0) + 15J_1(\omega_0) + 6J_2(2\omega_0)\} \qquad (3)$$

where $\chi_D = (eQeq_{zz}/h)$ is DQCC. The spectral densities $J_0(0)$, $J_1(\omega_0)$ and $J_2(2\omega_0)$ represent the Fourier transforms of the correlation functions characterizing the molecular motion in time. The orientational correlation functions can be computed within a jump-exchange model.[7, 18-21] The relevant expressions are given in the Supporting Material. In both cases, the temperature dependence of the correlation time $\tau_{ND}$ is expressed in terms of Arrhenius' law:

$$\tau_{ND} = \tau_{ND,\infty} \exp\left(\frac{E_a}{RT}\right) \qquad (4)$$

Temperature dependence of spin-relaxation times is presented in Figure 6. Relaxation time $(T_1)_D$ has a well-pronounced minimum at 220 K for the solid state II and at 263 K for the liquid state I (marked (i) and (iii)). Between 233 and 263 K $(T_1)_D$ and $(T_2)_D$ relaxation times for the isotropic signal do not coincide, therefore the extreme narrowing limit $\omega\tau \ll 1$ is not fulfilled at these temperatures. Above 263 K isotropic rotation of N–D bond governs the relaxation and its activation barrier can be determined from the slope of the relaxation curve.

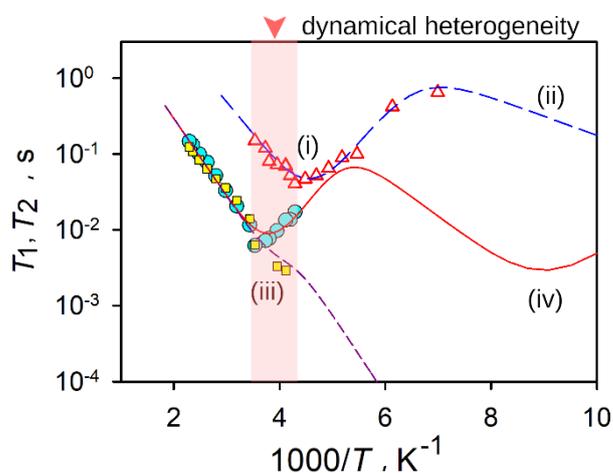

**Figure 6.** $^2$H NMR spin relaxation temperature dependences: experimental $T_1$ (△) for anisotropic state I; experimental $T_1$ (●) and $T_2$ (□) for isotropic mobile state II. Simulations are given in lines. All experimental data was measured at 61.4 MHz resonance frequency.

The virtue of the using N–D groups as a probe of molecular mobility is that there is no any internal rotation. Yet, the experimental results show, that the even in the solid state at two distinct anisotropic dynamic modes are present (Fig. 6). We assume that the N–D bond is involved into a restricted torsional librations ($k_{tor}$), responsible for the high temperature minimum of the $T_1$ relaxation curve (i in Fig. 6) and restricted librations in a cone ($k_{lib}$), governing the low temperature region (ii in Fig. 6). The torsion libration of the N–D group is related to the geometry of the hydrogen bond, with its orientation shifting between the oxygens of the anion. The additional soft librations represent the overall libration of the whole cation.

In the isotropic phase II we use the model previously developed for the hydrogen bond dynamics in [TEA][OTf][3] and [TEA][OMs][15] systems which includes the isotropic diffusion (iii in Fig. 6) of the whole cation and the restricted librations (iv in Fig. 6) of the N–D bond vector. The quadrupolar coupling constant used to fit the relaxation times were $\chi_{DI}$ = 173 kHz, obtained directly from the line shape analysis and $\chi_{DII}$ = 203 kHz estimated based on computational modes for the liquid state.[2] The angle of libration in the liquid state significantly rises from ±10° to ±42°. More restricted dynamics in the solid state can be rationalized by the existence of the double hydrogen bonds network. In both cases our dynamic models provide a perfect description of the experimental data. The model parameters are summarized in Table 1.

**Table 1.** Dynamical parameters of the [TEA][NTf$_2$] mobility in solid state I and liquid state II used for the simulation of spin relaxation curves.

|  | Solid State I | Liquid State II |
|---|---|---|
| $E_{tor}$, kJ mol$^{-1}$ | 17 | – |
| $k_{tor0}/(2\pi)$, Hz | 1x10$^{12}$ | – |
| $\phi_{tor}/\vartheta_{tor}$, deg | 47°/30° | – |
| $E_{lib}$, kJ mol$^{-1}$ | 5 | 10 |
| $k_{lib0}/(2\pi)$, Hz | 0.5x10$^{12}$ | 5x10$^{12}$ |
| $\vartheta_{lib}$, degrees | 10° | 35.5° |
| $E_{iso}$, kJ mol$^{-1}$ | – | 20 |
| $k_{iso0}/(2\pi)$, Hz | – | 9x10$^{11}$ |

The estimated accuracy for activation barriers is 10%.

For the preexponential factors – 50%.

Kinetic parameters of isotropic motion agree very well with parameters ($E_a$ = 21 kJ mol$^{-1}$ ; $k_0$ = 4.4×10$^{12}$ Hz) determined at different magnetic field ($\omega_0$ = 75.06 MHz),[22] while the field cycling relaxometry overestimates the barrier ($E_a$ = 30 kJ mol$^{-1}$, $k_0$ = 1.1x10$^{13}$ Hz). It is interesting to compare the mobility of [TEA][NTf$_2$] with the mobility of [TEA][OTf][3] and [TEA][OMs][15] reported earlier. The activation barrier of isotropic rotation in case of stronger anion [OMs]$^-$ is 25 kJ mol$^{-1}$ that is higher than 20 kJ mol$^{-1}$ for [OTf]$^-$ and [NTf$_2$]$^-$ based ILs. The libration angle in the liquid state II correlates with anion strength and drops down when the hydrogen bond becomes stronger ±23.5° < ±35° < ±42° for [OMs]$^-$, [OTf]$^-$ and [NTf$_2$]$^-$ correspondingly. At the same time the kinetic parameters of librational motion are insensitive to the anion type.

## 4. Conclusion

In present work the melting transition of [TEA][NTf$_2$] has been characterized by means of solid state $^2$H NMR spectroscopy. We show that the dynamical melting occurs through a dynamically heterogeneous phase stable between 223-277 K and the transition process is characterized by two standard molar enthalpy changes which characterize different stages of the melting. The spin relaxation analysis allowed determining geometry, rates and energetics of IL mobility in both solid and liquid state. Comparison of derived properties with previously studied [TEA][OTf] and [TEA][OMs] ILs that share cation but have anions of varying strength has been performed. Our results prove that the stronger hydrogen bonds between cation and anion lead to the lower enthalpy change between solid and liquid state, higher activation barrier of tumbling motion and lower amplitude of libration motion.

### Acknowledgement

This work has been supported by the Russian Science Foundation (grant № 21-13-00047).

### References

1.  S. Menne, J. Pires, M. Anouti and A. Balducci, *Electrochem. Commun.*, 2013, **31**, 39-41.
2.  A. E. Khudozhitkov, P. Stange, B. Golub, D. Paschek, A. G. Stepanov, D. I. Kolokolov and R. Ludwig, *Angew. Chem., Int. Ed.*, 2017, **56**, 14310-14314.
3.  A. E. Khudozhitkov, P. Stange, A. M. Bonsa, V. Overbeck, A. Appelhagen, A. G. Stepanov, D. I. Kolokolov, D. Paschek and R. Ludwig, *Chem. Commun.*, 2018, **54**, 3098-3101.
4.  K. Fumino, V. Fossog, K. Wittler, R. Hempelmann and R. Ludwig, *Angew. Chem., Int. Ed.*, 2013, **52**, 2368-2372.
5.  J. G. Powles and J. H. Strange, *Proc. Phys. Soc., London*, 1963, **82**, 6−15.

6. T. C. Farrar and E. D. Becker, *Pulse and Fourier Transform NMR. Introduction to Theory and Methods*, Academic Press, New York and London, 1971.
7. R. J. Wittebort, E. T. Olejniczak and R. G. Griffin, *J. Chem. Phys.*, 1987, **86**, 5411−5420.
8. V. Macho, L. Brombacher and H. W. Spiess, *Appl. Magn. Reson.*, 2001, **20**, 405-432.
9. R. Ludwig, F. Weinhold and T. C. Farrar, *J. Chem. Phys.*, 1995, **103**, 6941-6950.
10. R. Ludwig, *Chem Phys*, 1995, **195**, 329-337.
11. R. Ludwig, D. S. Gill and M. D. Zeidler, *Z. Naturfors. Sect. A-J. Phys. Sci.*, 1991, **46**, 89-94.
12. R. Ludwig and M. D. Zeidler, *Molecular Physics*, 1994, **82**, 313-323.
13. R. Ludwig and M. D. Zeidler, *Zeitschrift Fur Physikalische Chemie-International Journal of Research in Physical Chemistry & Chemical Physics*, 1995, **189**, 19-27.
14. M. Susan, A. Noda, S. Mitsushima and M. Watanabe, *Chem. Commun.*, 2003, DOI: 10.1039/b300959a, 938-939.
15. A. E. Khudozhitkov, V. Overbeck, P. Stange, A. Strate, D. Zaitsau, A. Appelhagen, D. Michalik, A. G. Stepanov, D. I. Kolokolov, D. Paschek and R. Ludwig, *Phys. Chem. Chem. Phys.*, 2019, **21**, 25597-25605.
16. G. E. Pake, *J. Chem. Phys.*, 1948, **16**, 327-336.
17. A. Abragam, *The Principles of Nuclear Magnetism*, Oxford University Press, Oxford, UK, 1961.
18. H. W. Spiess, in *NMR Basic Principles and Progress*, eds. P. Diehl, E. Fluck and R. Kosfeld, Springer-Verlag, New York, 1978, vol. 15, p. 55−214.
19. R. J. Wittebort and A. Szabo, *J. Chem. Phys.*, 1978, **69**, 1722–1736.
20. G. Lipari and A. Szabo, *Biophys. J.*, 1980, **30**, 489−506.
21. L. J. Schwartz, E. Meirovitch, J. A. Ripmeester and J. H. Freed, *J. Phys. Chem.*, 1983, **87**, 4453−4467.
22. V. Overbeck, B. Golub, H. Schroder, A. Appelhagen, D. Paschek, K. Neymeyr and R. Ludwig, *J. Mol. Liq.*, 2020, **319**, 7.